\documentclass[twocolumn, prl, amssymb, superscriptaddress, aps, showpacs,
preprintnumbers,amsmath,floatfix]{revtex4}
\usepackage{multirow}
\usepackage{isotope}
\usepackage{supertabular}
\usepackage{graphicx,amsfonts,color,epsfig}
\usepackage{amsmath}
\usepackage{amssymb}
\usepackage{url}

\begin{document}

\title{Reduced transition probabilities for the $\gamma$-decay
of the 7.8 eV isomer in $^{229}$Th}

\author{Nikolay \surname{Minkov}}
\email{nminkov@inrne.bas.bg} \affiliation{Institute of Nuclear Research and
Nuclear Energy, Bulgarian Academy of Sciences, Tzarigrad Road 72, BG-1784
Sofia, Bulgaria} \affiliation{Max-Planck-Institut f\"ur Kernphysik,
Saupfercheckweg 1, D-69117 Heidelberg, Germany}

\author{Adriana P\'alffy}
\email{Palffy@mpi-hd.mpg.de}
\affiliation{Max-Planck-Institut f\"ur Kernphysik, Saupfercheckweg 1,
D-69117 Heidelberg, Germany}


\date{\today}

\begin{abstract}

The reduced magnetic dipole  and electric quadrupole  transition probabilities for the radiative
decay of the $^{229}$Th 7.8~eV isomer to the ground state are predicted within a detailed
nuclear-structure model approach. We show that the presence and decay of this isomer can
only be accounted for by the Coriolis mixing emerging from a remarkably fine interplay
between the coherent quadrupole-octupole motion of the nuclear core and the single-nucleon
motion within a reflection-asymmetric deformed potential. We find that the magnetic dipole
transition probability which determines the radiative lifetime of the isomer is considerably smaller than presently estimated. The so-far disregarded electric quadrupole
component may have non-negligible contributions to the internal conversion channel. These
findings support new directions in the experimental search of the $^{229}$Th transition
frequency for the development of a future nuclear frequency standard.

\end{abstract}

\pacs{
23.35.+g  
23.20.Lv, 
21.60.Ev, 
21.60.Cs, 
06.30.Ft 
}



\maketitle

{\it Introduction.}
From the nuclear structure point of view, the $^{229}$Th actinide isotope is
a typical representative of a heavy nucleus with pronounced collectivity and
possible presence of octupole (reflection-asymmetric) deformation
\cite{BN96}. The increasing interest that $^{229}$Th has received in the last
decade beyond the nuclear physics community is due to an exceptional
phenomenon: the presence of a $3/2^{+}$ isomeric, i.e., long-lived, state
identified at energy of approx. $7.8$ eV, which appears as an almost
degenerate counterpart of the $5/2^{+}$ ground state \cite{Beck0709}. This
extremely small energy (the smallest known up to date in nuclear spectra)
renders for the first time a nuclear transition accessible to vacuum
ultra-violet (VUV) lasers. Given the very narrow linewidth of the transition
and the high robustness of nuclei to external perturbations
\cite{clock_campbell_2012}, the isomeric state has been proposed for novel
applications such as a nuclear frequency standard with unprecedented accuracy
\cite{clock_peik_2003,clock_campbell_2012,clock_peik_2015} or a nuclear
laser~\cite{nucl_laser}.

The first direct observation of the low-energy isomeric state decay of $^{229}$Th  via internal
conversion (IC) has been reported  recently \cite{Wense16}. The most recent energy value of
$7.8 \pm 0.5$ eV could be determined only indirectly in a calorimetric measurement by
subtraction of x-ray energy differences between neighboring levels \cite{Beck0709}.
So far, a direct excitation of the isomeric state has not been achieved \cite{Jeet15,yamaguchi},
and the radiative lifetime of the transition to the ground state could only be estimated
theoretically. A nuclear structure theory prediction of the isomer energy on the eV level of
accuracy is beyond reach. On the other hand, predictions for the reduced $3/2^{+}\rightarrow 5/2^{+}$ transition
probabilities have been attempted on the basis of branching ratios (Alaga rules \cite{Alaga55})
from the observed decays of neighboring levels \cite{Dyk98,Tkalya15}, as well as  within the quasiparticle-plus-phonon model (QPM)
\cite{Sol76} in Refs.~\cite{Gulda02,Ruch06}. Due to the very small energy it is assumed that the magnetic dipole
$(M1)$ dominates over the electric quadrupole $(E2)$ for the  transition from the
isomeric $3/2^{+}$ to the ground $5/2^{+}$ state.

So far, the unusually low isomer energy and its decay properties are not well understood from the nuclear
structure point of view.  In this Letter, we report the elaboration and
first-time application of a sophisticated model approach which incorporates the
shape-dynamic properties together with the intrinsic structure characteristics typical for the
actinide nuclei. The formalism includes a description of the collective quadrupole-octupole
vibration-rotation motion of the nucleus which in the particular case of odd-mass nuclei is
coupled to the motion of the single (odd) nucleon within a reflection-asymmetric
deformed-shell model with pairing correlations and fully microscopic treatment of the
Coriolis interaction.  The model calculation determines the energy and the radiative decay
properties  of this isomer as integral parts of the entire low-lying positive- and negative-parity
spectrum and transition probabilities observed in $^{229}$Th. On this basis, we show that 
the low-lying isomer energy emerges as the consequence of a very fine interplay between
the rotation-vibration degrees of freedom and the coupling to the motion of the unpaired neutron.
The predicted transition probability value   $B(M1)=0.007$ 
Weisskopf units (W.u.) lies about one order of magnitude lower than the value of 0.048 W.u. deduced in Ref.~\cite{Dyk98} 
and by a factor of two lower than the value of 0.014 W.u. obtained from the QPM model \cite{Ruch06}. This might offer
an explanation for recent experimental difficulties to observe the radiative decay of the isomer \cite{Jeet15,yamaguchi,LarsThesis2016}.
The electric quadrupole value $B(E2)=27$ W.u. is by a factor of approx.~2 smaller than the only other available prediction in
Ref.~\cite{Ruch06}. Despite the recent claims in Ref.~\cite{Tkalya15} that the electric quadrupole channel is negligible
for both radiative and IC decays, for excited electronic state configurations, the predicted values render the
 contribution of the $E2$ transition for IC  of equal magnitude with
the $M1$ one \cite{Pavlo2017}. These findings support experimental efforts under way that
focus on the IC from the isomeric state \cite{Seiferle2017} as opposed to experiments  involving so far mostly 
 the radiative decay of the isomer
\cite{Jeet15,yamaguchi,LarsThesis2016}.

{\it Theoretical approach.}
The complex nuclear structure of the $^{229}$Th isotope is governed by the fine interplay
between the collective quadruple-octupole (QO) vibration-rotation motion of the nucleus, the
single-particle (s.p.) motion of the odd, unpaired nucleon and the Coriolis interaction between
the latter and the nuclear core. The collective motion is described through the so-called
coherent QO mode (CQOM) giving raise to the quasi parity-doublet structure of the spectrum
\cite{b2b3mod,b2b3odd}, whereas the s.p.~one is determined by the deformed shell model
(DSM) with reflection-asymmetric Woods-Saxon (WS) potential \cite{qocsmod} and pairing
correlations of Bardeen-Cooper-Schrieffer (BCS) type included as in Ref.~\cite{WM10}. The
Coriolis interaction between CQOM and the odd nucleon is considered following
Refs.~\cite{MDSS09_10,NM13}.

The Hamiltonian of QO vibrations and rotations coupled to the s.p. motion
with Coriolis interaction and pairing correlations can be written in the form
\begin{eqnarray}
H=H_{\mbox{\scriptsize s.p.}}+H_{\mbox{\scriptsize pair}}+
H_{\mbox{\scriptsize qo}}+H_{\mbox{\scriptsize Coriol}}\, .
\label{Htotal}
\end{eqnarray}
Here $H_{\mbox{\scriptsize s.p.}}$ is the s.p. Hamiltonian with the WS potential for axial
quadrupole, octupole and higher multipolarity deformations \cite{qocsmod} providing the
s.p. energies $E^{K}_{\mbox{\scriptsize sp}}$ with given value of the projection $K$ of the
total and s.p. angular momentum operators $\hat{I}$ and $\hat{j}$ on the intrinsic symmetry
axis. $H_{\mbox{\scriptsize pair}}$ is the standard BCS pairing Hamiltonian \cite{RS80}.
Together they yield the quasi-particle (q.p.) spectrum $\epsilon^{K}_{\mbox{\scriptsize
qp}}$ as shown in Ref.~\cite{WM10}. $H_{\mbox{\scriptsize qo}}$ represents oscillations of
the even--even core with respect to the quadrupole ($\beta_2$) and octupole ($\beta_3$) axial
deformation variables mixed through a centrifugal (rotation-vibration) interaction
\cite{b2b3odd}. $H_{\mbox{\scriptsize Coriol}}$ involves the Coriolis interaction between
the even-even core and the unpaired nucleon (see Eq.~(3) in \cite{b2b3odd}). It is treated as a
perturbation with respect to the remaining part of (\ref{Htotal}) and then incorporated into the
QO potential of $H_{\mbox{\scriptsize qo}}$ defined for given angular momentum $I$,
parity $\pi$  and s.p. band-head projection $K_{b}$ which leads to a joint term \cite{NM13}
\begin{eqnarray}
H_{\mbox{\scriptsize qo}}^{IK_{b}}
&=& -\frac{\hbar^2}{2B_2}\frac{\partial^2}{\partial\beta_2^2}
-\frac{\hbar^2}{2B_3}\frac{\partial^2}{\partial\beta_3^2}+
\frac{1}{2}C_2{\beta_2}^{2}+
\frac{1}{2}C_3{\beta_3}^{2} \nonumber \\
&+& \frac{\widetilde{X}(I^{\pi},K_{b})}
{d_2\beta_2^2+d_3\beta_3^2}\, . \label{HqoK}
\end{eqnarray}
Here,  $B_2$ $(B_3)$, $C_2$ $(C_3)$ and $d_2$ ($d_3$) are quadrupole (octupole) mass,
stiffness and inertia parameters, respectively, and $\widetilde{X}(I^{\pi},K_{b})$ determines
the centrifugal term in which the Coriolis mixing is taken into account (see Supplementary
Material \cite{Supplmat} for details).

The Coriolis perturbed wave function for the QO spectrum built on a q.p. state with
$K=K_{b}$ and parity $\pi^{b}$ corresponding to the Hamiltonian (\ref{Htotal}) is obtained in
the first order of perturbation theory from the QO core plus particle wave function \cite{NM13} 
(see Supplementary Material \cite{Supplmat}). The corresponding expression for the energy has the form
\begin{equation}
E_{nk}^{\mbox{\scriptsize tot}}(I^{\pi} ,K_{b})
=\epsilon^{K_{b}}_{\mbox{\scriptsize qp}}
+ \hbar\omega \left[ 2n+1+\sqrt{k^2+b\widetilde{X}(I^{\pi},K_{b})}\right],
\label{enspect1}
\end{equation}
where $b=2B/(\hbar^2 d)$ denotes the reduced inertia parameter and $n=0,1,2...$
and $k=1,2,3,...$ stand for the radial and angular QO oscillation quantum numbers,
respectively, with $k$ odd (even) for the even (odd) parity states of the core 
\cite{MDSSL12}.  The levels of the total QO core plus particle system, determined by a particular
$n$ and $k^{(+)}$ ($k^{(-)}$) for the states with given $I^{\pi=+}$ ($I^{\pi=-}$) form a split
(quasi) parity-doublet \cite{MDDSLS13}. Furthermore, 
$\omega=\sqrt{C_2/B_2}=\sqrt{C_3/B_3}\equiv \sqrt{C/B}$ stands for the frequency of the
coherent QO oscillations  \cite{b2b3mod,b2b3odd}.

Having the Coriolis perturbed wave function we were able to calculate  the reduced
probabilities $B(E1)$, $B(E2)$, $B(E3)$ and $B(M1)$  for transitions between initial (i) and
final (f) states with energies given by Eq.~(\ref{enspect1}) (see Supplementary Material \cite{Supplmat} for the
explicit expressions). We note that the reduced transition probability expression contains
first-order and second-order $K$-mixing effects. First-order mixing terms practically
contribute with non-zero values only in the cases $K_{i/f}=K_{\nu}=1/2$, i.e., when a
$K_{i/f}=1/2$ band-head state is mixed with another $K_{\nu}=1/2$ state present in the
considered range of admixing orbitals above the Fermi level. A second order mixing effect
connects states with $\Delta K=1,2$ and allows different combinations of $|K_i-K_f|\leq 2$
which provide respective non-zero contribution of the Coriolis mixing to the transition
probability. Therefore,  non-zero transition probabilities between states with different
$K$-values are rendered possible. Thus, unlike stated in previous works
\cite{Dyk98,Tkalya15}, in our model {\em it is the Coriolis interaction which allows the
otherwise forbidden $M1$ and $E2$ inter-band isomeric transition to occur}. This allows us
to examine in detail the {\em capability of such a complete nuclear structure mechanism to
provide a prediction for the isomer decay in} $^{229}$Th based on the interaction between
collective and s.p. degrees of freedom in the nucleus.

\begin{figure}[t]
\centering
\centerline{\includegraphics[width=0.5\textwidth]{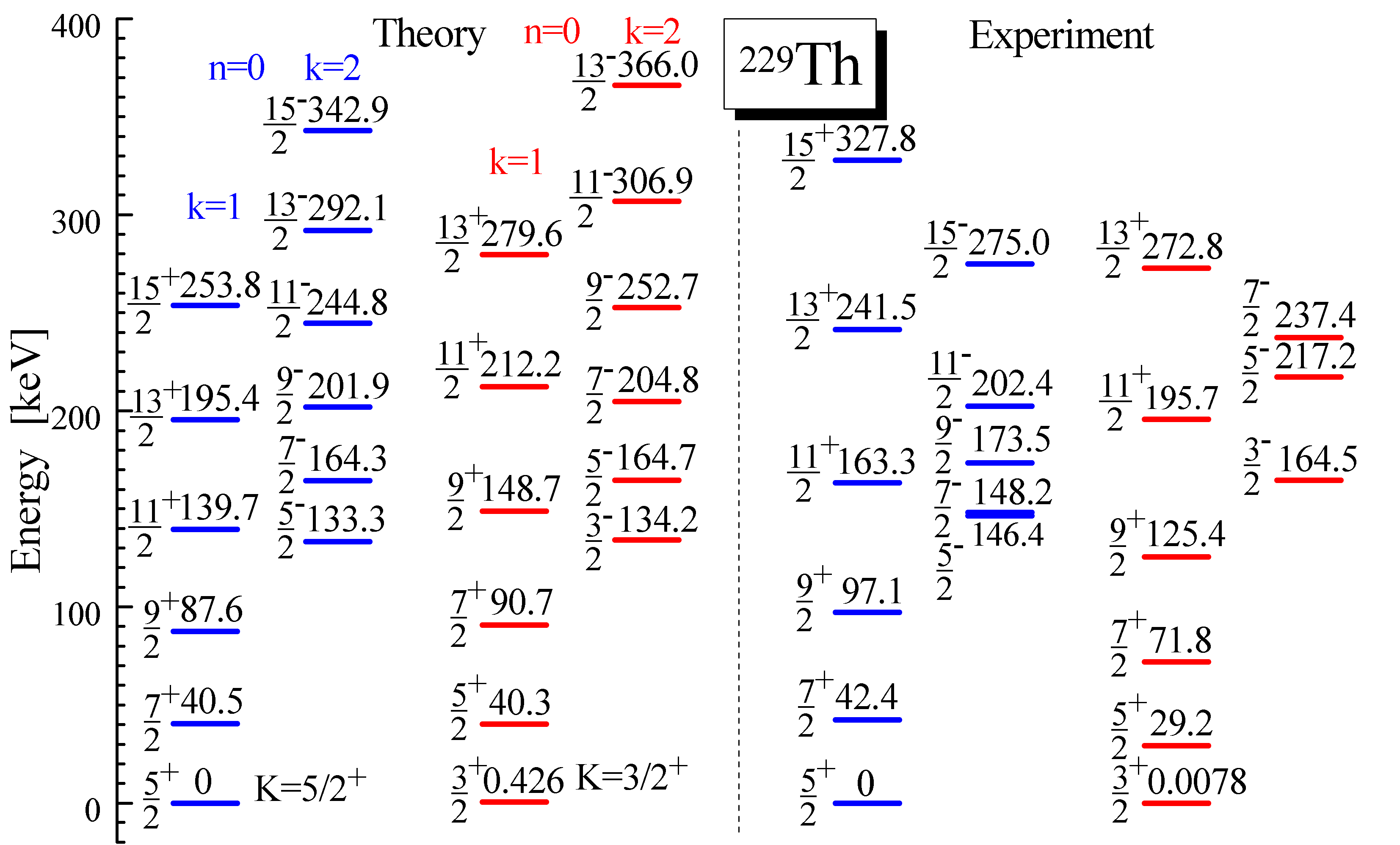}}
\caption[]{Theoretical and experimental quasi parity-doublet levels of
$^{229}$Th. The CQOM model parameters used are $\omega=0.20$ MeV/$\hbar$,
$b=0.28$ $\hbar^{-2}$, $d_0=18$ $\hbar^{2}$, $c=79$, $p=1$ and $A=0.158$ keV. See text for further explanations.
Experimental data from Ref.~\cite{ensdf}.}
\label{fig:th229}
\end{figure}

{\it Numerical results.}
We applied the  above CQOM-DSM approach to the low-lying part of the
experimental $^{229}$Th spectrum \cite{ensdf} including positive- and
negative-parity levels with energy below 400 keV as shown in
Fig.~\ref{fig:th229}. We suggest that this is enough to constrain the CQOM
model parameters  as to provide reliable predictions for the unknown $B(E2)$
and $B(M1)$ transition probabilities between the isomeric $3/2^{+}$ and the
$5/2^{+}$ ground state. This spectrum is interpreted according to CQOM as
consisting of two parity quasi-doublets as follows.

The lower, yrast (yr) quasi-doublet, is built on the $K_{b^{\mbox{\scriptsize yr}}}=5/2^{+}$
ground state corresponding to the 5/2[633] Nilsson s.p. orbital. The set of positive- and
negative-parity levels is interpreted as collective rotations and coherent vibrations of the QO
deformed core-plus-particle system. The parity splitting is due to the QO vibration mode
characterized by the lowest value of the radial-oscillation quantum number $n=0$, the lowest
possible angular-oscillation number $k_{\mbox{\scriptsize yr}}^{(+)}=1$ for the
positive-parity sequence and one of the few lowest possible $k_{\mbox{\scriptsize
yr}}^{(-)}=2,4,6$ values for the negative-parity states.

The upper, excited (ex), i.e., non-yrast, quasi parity-doublet is built on the isomeric
$K_{b^{\mbox{\scriptsize ex}}}=3/2^{+}$ state corresponding to the 3/2[631] Nilsson s.p.
orbital. For other nuclei, the first non-yrast quasi-doublet is usually associated with the higher
radial-vibration mode, $n=1$ \cite{MDDSLS13}. However, having in mind the practically full
degeneracy of the pair of $5/2^{+}$ and $3/2^{+}$ states, here we consider that this doublet
also corresponds to the collective QO mode characterized by $n=0$ with the lowest
$k_{\mbox{\scriptsize ex}}^{(+)}=1$ and one of the lowest possible $k_{\mbox{\scriptsize
ex}}^{(-)}=2,4,6,8$.

Thus, we have two similar collective quasi-doublet structures with $n=0$,
$k_{\mbox{\scriptsize yr}}^{(+)}=k_{\mbox{\scriptsize ex}}^{(+)}=1$, built on the two
quasi-degenerate s.p. states. This theoretical supposition is supported by the observation that
both the yrast and non-yrast experimental positive-parity sequences have very similar level
spacings up to $I=13/2^{+}$ and only differ by a mutual shift. The latter can be explained by
the different $K_{b}$-values in the centrifugal term $\widetilde{X}(I^{\pi},K_{b})$ in Eq.~(\ref{HqoK})
leading to a relative down-shift of the $5/2^{+}$ sequence with respect to $3/2^{+}$
\cite{Supplmat}. The experimental negative-parity sequences in both quasi-doublets, though
not looking as similar as the positive-parity ones, are also reasonably placed in this scheme
suggesting equal or close values of $k_{\mbox{\scriptsize yr}}^{(-)}$ and
$k_{\mbox{\scriptsize ex}}^{(-)}$.

The calculations were made for several combinations of 
$k_{\mbox{\scriptsize yr}}^{(-)}$ and $k_{\mbox{\scriptsize ex}}^{(-)}$
values providing respective sets of adjusted CQOM model parameters as detailed in the following. We consider the
lowest values $k_{\mbox{\scriptsize yr}}^{(-)}=k_{\mbox{\scriptsize
ex}}^{(-)}=2$  as the basic set providing the lowest-energy QO vibration
mode. On the other hand the calculation with higher $k^{(-)}$-values allows
us to examine the stability of the obtained predictions against different
model conditions.

We have first determined the quadrupole ($\beta_2$) and octupole ($\beta_3$)
deformations in DSM by requiring that the orbitals 5/2[633] and 3/2[631]
appear in the  neutron s.p. spectrum as the last occupied and first not
occupied orbitals, respectively, and the spacing between them is as small as
possible. By varying $\beta_2$ between the experimental values 0.2301(39) and
0.2441(15) known for the neighboring even-even nuclei $^{228}$Th and
$^{230}$Th, respectively \cite{K02}, and changing additionally $\beta_3$ we
obtained $\beta_2=0.240$ and $\beta_3=0.115$. It is important to note that
we found the correct placing and mutual spacing of both orbitals {\em at
non-zero octupole deformation}.

In the following, starting from the set of $k_{\mbox{\scriptsize yr}}^{(-)}=k_{\mbox{\scriptsize
ex}}^{(-)}=2$ we performed fits of the  six CQOM model parameters for fixed BCS pairing
constants (see Supplementary Material \cite{Supplmat} for details). Apart from $\omega$ and $b$, further model parameters are
$d_0$, which fixes the QO potential origin \cite{b2b3mod,b2b3odd} and  together with $\omega$ and $b$ determines the energy levels; $c=(B/\hbar ) \omega$, the
 reduced oscillator frequency, and $p=\sqrt{(d_{2}+d_{3})/(2d_{2})}$ which gives the relative contribution of the quadrupole mode
\cite{MDSSL12}, both additionally determining the transition probabilities. The sixth parameter is the Coriolis
mixing strength $A$ which determines both energies and transitions.  This procedure allowed us to obtain the theoretical
$3/2^{+}$ isomer energy value as small as 0.4 keV. The theoretical energy levels are shown in
Fig.~\ref{fig:th229} in comparison with the respective experimental data. The root mean
square (rms) deviation of the predicted energy levels from the experimental ones is
rms$_{\mbox{\scriptsize yr}}=39.9$ keV for the yrast-based band and
rms$_{\mbox{\scriptsize ex}}=26$ keV for the isomer-based band, with the total deviation
being rms$_{\mbox{\scriptsize tot}}=34$ keV. In this calculation the predicted $B(E2)$ and
$B(M1)$ values for the $3/2^{+}$ isomer decay are 27.04 W.u. and 0.0076 W.u. respectively.
The theoretical (Th1) and experimental \cite{nndc_gam} values for the available $B(E2)$ and
$B(M1)$ transition probabilities are compared in Table~\ref{tab:trans}.

Though from a nuclear physics point of view  0.4 keV is a very good approximation of the
experimental 0.0078 keV, faced with atomic physics accuracy standards  a refinement is
desirable. Therefore, we implemented one further step by slightly varying the CQOM
parameters values given in the caption of Fig.~\ref{fig:th229} to individually adjust the
theoretical $3/2^{+}$ isomer energy to the exact experimental value. As might be
expected, this led to a slight overall deterioration of the predictions for the other energy levels,
with an increase of rms$_{\mbox{\scriptsize tot}}$ by 1 keV. At the same time the predicted
$B(E2)$ and $B(M1)$ values slightly changed to 23.05 W.u. and 0.0061 W.u.,
respectively.  To see the effect of this $3/2^{+}_{\mbox{\scriptsize ex}}$ adjustment on the
other described transition probabilities we present the respective theoretical values (Th2) in
Table~\ref{tab:trans}. The overall change in the obtained transition probabilities is small.
This refinement illustrates the sensitivity of the obtained description and predictions to the
model parameters in reference to the accuracies inherent for the atomic physics quantities.
Also, it shows how the collective QO-oscillation mode affects the band-head energy. Further
checks were performed keeping the same BCS conditions fixed for few more combinations of
higher $k_{\mbox{\scriptsize yr}}^{(-)}$ and $k_{\mbox{\scriptsize ex}}^{(-)}$ values
without tuning the $3/2^{+}$ isomer energy (see Supplementary Material \cite{Supplmat}). The predicted isomer $B(E2)$ and $B(M1)$ values do not
change very much and vary in the limits between 20 and 30 W.u. for $B(E2)$ and 0.006 and
0.008 W.u. for the $B(M1)$ transition probabilities, confirming the prediction stability of the
model.

\begin{table}[htbp]
\caption{Theoretical and where available experimental $B(E2)$ and $B(M1)$
transition values for the low-lying spectrum of $^{229}$Th. Results are given
for two sets of model parameters, denoted as Th1 (corresponding to
Fig.~\ref{fig:th229}) and Th2 presented in the text. Experimental data from
Ref.~\cite{nndc_gam}.}
\begin{center}
{\small
\begin{tabular}{cccc}
\hline\hline
Type& Transition & Th1[Th2] (W.u.) & Exp (Err) (W.u.)\\
\hline \hline
E2&$7/2^{+}_{\mbox{\scriptsize yr}}$ $\rightarrow$ $5/2^{+}_{\mbox{\scriptsize yr}}$ & 252 [267] & 300 ($\pm$16) \\
E2&$9/2^{+}_{\mbox{\scriptsize yr}}$ $\rightarrow$ $5/2^{+}_{\mbox{\scriptsize yr}}$ & 82 [85]  & 65  ($\pm$7)\\
E2&$9/2^{+}_{\mbox{\scriptsize yr}}$ $\rightarrow$ $7/2^{+}_{\mbox{\scriptsize yr}}$ & 213 [224] & 170  ($\pm$30)\\
E2&$9/2^{+}_{\mbox{\scriptsize yr}}$ $\rightarrow$ $5/2^{+}_{\mbox{\scriptsize ex}}$ & 19.98 [17.37] & 6.2 ($\pm$0.8)\\
E2&$\mathbf{3/2^{+}_{\mbox{\scriptsize ex}}\ \rightarrow \ 5/2^{+}_{\mbox{\scriptsize yr}}}$ & $\mathbf{27.04}$ $\mathbf{[23.05]}$ & --  \\
M1&$7/2^{+}_{\mbox{\scriptsize yr}}$ $\rightarrow$ $5/2^{+}_{\mbox{\scriptsize yr}}$ & 0.0093 [0.0085] & 0.0110 ($\pm$0.0040) \\
M1&$9/2^{+}_{\mbox{\scriptsize yr}}$ $\rightarrow$ $7/2^{+}_{\mbox{\scriptsize yr}}$ & 0.0178 [0.0157] & 0.0076 ($\pm$0.0012) \\
M1&$9/2^{+}_{\mbox{\scriptsize yr}}$ $\rightarrow$ $7/2^{+}_{\mbox{\scriptsize ex}}$ & 0.0151 [0.0130] & 0.0117 ($\pm$0.0014) \\
M1&$\mathbf{3/2^{+}_{\mbox{\scriptsize ex}}\ \rightarrow\ 5/2^{+}_{\mbox{\scriptsize yr}}}$ & $\mathbf{0.0076}$  $\mathbf{[0.0061]}$   & --  \\
\hline \hline
\end{tabular}
\label{tab:trans}
}
\end{center}
\end{table}

{\it Discussion.} The model description of the low-lying positive- and negative-parity levels
of $^{229}$Th shown in Fig.~\ref{fig:th229} provides a reasonable interpretation of the
$3/2^{+}$ isomeric state as a part of the quasi-doublet structure stemming from the general
QO rotation-vibration degrees of freedom of the total system and the coupling  between the
even-even core and the motion of the unpaired neutron. Our detailed model calculations
suggest that the extremely small isomer energy  may be the result of a very fine interplay
between all involved degrees of freedom.  We emphasize the crucial importance of the
Coriolis interaction for explaining the presence of an $M1$ transition between the yrast and
non-yrast band, contradicting statements in the literature on the weakness of the Coriolis
mixing \cite{Dyk98,Tkalya15}. At the same time our result corroborates with the role of the
Coriolis mixing assumed in the QPM calculations \cite{Gulda02,Ruch06}. We note here that 
the present model is essentially different from the QPM application. In the latter
the mean field is supposed to be reflection symmetric, whereas the octupole mode is included
as a phonon admixture to the q.p. excitation. In the present approach the quadrupole and
octupole modes are treated on the same footing both in DSM and in the collective CQOM part,
reflecting the present understanding about the stronger role of the octupole
deformation in the shape dynamics of actinide nuclei. Furthermore, in the
CQOM+DSM model the Coriolis term is not diagonalized numerically but treated explicitly within
the perturbation theory providing a possibility to follow in detail the role of the $K$-mixing in
the mechanism of isomer decay. The limits of validity of this
perturbative approach were examined in Ref.~\cite{NM13} for the quasi parity-doublet spectra of
$^{223}$Ra and $^{221}$Fr, where a much stronger Coriolis interaction is observed.


Inspection of our results in Table~\ref{tab:trans} shows that the experimental data 
is reproduced quite well with some entries like the magnetic intraband
$7/2^{+}_{\mbox{\scriptsize yr}}\rightarrow 5/2^{+}_{\mbox{\scriptsize yr}}$ theoretical value being
 within the experimental uncertainties. A larger discrepancy is observed for the
$9/2^{+}_{\mbox{\scriptsize yr}}\rightarrow 7/2^{+}_{\mbox{\scriptsize yr}}$ $M1$ transition.
This might be related to the larger discrepancy (about 10 keV) of the predicted energy for the
$9/2^{+}_{\mbox{\scriptsize yr}}$ level  compared to the one ($\sim 2$ keV) of
$7/2^{+}_{\mbox{\scriptsize yr}}$ (see Fig.~\ref{fig:th229}). We note that this particular
transition probability is overestimated also by QPM and different experimental values are available at present \cite{Ruch06,nndc_gam}
calling for further measurements. The interband $9/2^{+}_{\mbox{\scriptsize yr}}\rightarrow
7/2^{+}_{\mbox{\scriptsize ex}}$  $M1$ transition in Table~\ref{tab:trans} shows a better
agreement with the experiment, and our $B(E2; 9/2^{+}_{\mbox{\scriptsize
yr}}\rightarrow 5/2^{+}_{\mbox{\scriptsize ex}})$ values of 19.98 and 17.37 W.u. though
overestimating the ENSDF value of 6.2 ($\pm 0.8$) W.u., are both within the error bar of the
experimental value of 19.2 ($\pm 4.8$) W.u. reported in \cite{Ruch06}. Thus, having in mind
the possible uncertainties in the experimental data we may conclude that the present
calculations provide a reasonable overall description of the electromagnetic transition
properties of $^{229}$Th. In addition we note that CQOM provides plausible values for
the intrinsic quadrupole moment within the yrast and non-yrast sequences varying in the
range of $780-960$ fm$^2$ for $5/2^{+} \leq I \leq 15/2^{+}$, with the lower value being
close to the one of $750$ fm$^2$ considered in Ref.~\cite{Ruch06}.

The present formalism allows to study the fine interplay between all
involved dynamic modes, providing a way to exactly tune the isomer energy without essential
loss of overall accuracy through the six CQOM parameters. In order to check the reliability of the predicted decay
probabilities and the predictive capability of the model as a whole we have performed  
a similar calculation for  the neighboring odd-mass
isotope $^{231}$Th using the six parameter values obtained from the $^{229}$Th spectrum. 
We considered the yrast quasi parity-doublet band in $^{231}$Th together with two $B(E1)$ transition values available from experimental
data \cite{nndc_gam}. The only additional refinements we made were the
(necessary) use of an effective $E1$ charge of 0.43$e$, with $e$ the proton charge, and a slight shift of $\beta_2$ in DSM from
0.24 to $\beta_2=0.248$ to obtain the last occupied orbital with $K^{\pi}=5/2^{+}$. We found
quite a reasonable description of both energies and $B(E1)$ transition probabilities of
$^{231}$Th with the energy rms value being $26$ keV. Thus, we may conclude that our model
parameters demonstrate rather stable behavior among different isotopes in conjunction with the concept of
``sloppy'' parameter fits discussed in Ref.~\cite{NV16}. 

This analysis and the good
reproduction of the available values for intraband and interband transition rates suggest that
 the experimental transition probabilities for
the $3/2^{+}$-isomer decay in $^{229}$Th will be found in the limits of $B(E2)$=20--30
W.u. and $B(M1)$=0.006--0.008 W.u. The fact that the $B(M1)$ value is significantly smaller than the
theoretical predictions up to date  \cite{Dyk98,Ruch06,Tkalya15}  is in
agreement with and could be an explanation for the lack of experimental success in driving or
observing the radiative channel for the isomeric transition
\cite{Jeet15,yamaguchi,LarsThesis2016}. At the same time the suggested range for the
transition probabilities could serve as a clearly determined accuracy target for further
experiments. Furthermore, we note that the $E2$ channel, thus far disregarded, can play a role
for IC, especially when considering other electronic orbitals than the neutral Th atomic
ground state. Scaling the IC rates obtained in Ref.~\cite{Pavlo2017} using the previously
available larger value $B(M1)=0.048$ W.u., it turns out that the $E2$ component could even
be dominant over the $M1$ one for conversion involving the $7p$, $6d$ or $5f$ electronic
orbitals. Our new results on the $B(E2)$ and $B(M1)$ transition probabilities therefore
support experimental efforts aiming at the determination of the isomeric decay properties via
the observation of IC electrons.

This work is supported by the DFG and by the BNSF under contract DFNI-E02/6. AP
gratefully acknowledges funding by the EU FET-Open project 664732.


\end{document}